\documentclass[12pt]{prop}
\usepackage{pstricks}
\usepackage{apjfonts}
\usepackage{graphicx}
\usepackage{amsmath}
\renewcommand{\textfloatsep}{4mm}

\def\fsz{\footnotesize}

\def\bi{\bfseries\itshape}
\def\bs{\bfseries\sffamily}
\def\bsi{\bfseries\sffamily\itshape}

\renewcommand{\arraystretch}{1.5}

\renewcommand{\tabcolsep}{.7cm}

\def\msun       {$M_{\odot}$}
\def\deg        {$^{\circ}$}

\def\as         {$^{\prime\prime}$}
\def\chandra    {\emph{Chandra}}
\def\xmm        {\emph{XMM}}
\def\xmmnewt    {\emph{XMM-Newton}}
\def\swift      {\emph{Swift}}
\def\hitomi     {\emph{Hitomi}}
\def\xrism     {\emph{XRISM}}
\def\suzaku     {\emph{Suzaku}}
\def\nustar    {\emph{NuSTAR}}
\def\nicer     {\emph{NICER}}
\def\athena    {\emph{Athena}}
\def\lynx    {\emph{Lynx}}
\def\axis    {\emph{AXIS}}
\def\calx     {\emph{Cal \mbox{X-1}}}

\begin{document}
%\singlespace
\sloppy

\title{\sf\LARGE {\em Cal X-1}: an absolute in-orbit calibrator for current
  and future X-ray observatories}

\maketitle
\thispagestyle{empty}

\begin{center}

\vspace*{-1cm}
Project White Paper for Astro-2020 Decadal Survey

\vspace{2mm}
Type of activity: Space-Based Project 

\vspace*{1cm}
\begin{minipage}{16cm}
\centering {\em Authors}: 
Keith Jahoda$^{1,5,*}$, Maxim Markevitch$^1$, Takashi Okajima$^1$, Joanne
Hill-Kittle$^1$, Neerav Shah$^1$, Denis Bergeron$^2$, Andrew
Holland$^3$, Paul Plucinsky$^{4,5}$, Dan Schwarz$^{4,5}$

\vspace*{3mm} {\fsz $^1$NASA Goddard ~~$^2$NIST ~~$^3$Open University, UK
  ~~$^4$Harvard-Smithsonian CfA \\ ~~$^5$IACHEC ~~$^*$keith.jahoda@nasa.gov}

\vspace*{5mm} {\em Endorsed by}: Belinda Wilkes (\chandra\ X-ray Center
Director), \\ Norbert Schartel ({\em XMM-Newton}\/ Project Scientist), \\
Matteo 
Guainazzi (\athena\ Study Scientist, \xrism\ ESA Project Scientist)

\vspace*{1cm}
{\small July 10, 2019}

\end{minipage}
\end{center}

%%%%%%%%%%%%%%%%%%%%%%%%%%%%%%%%%%%%%%%%%%%%%%%%%%%%%%%%%%%%%%%%%%%%%%%%%%
\begin{figure*}[b]
\centering
\pspicture(0,0)(15,6.4)

\rput[bl]{0}(1.3,0.){%
\includegraphics[angle=0,height=7cm,bb=73 391 368 718,clip]%
{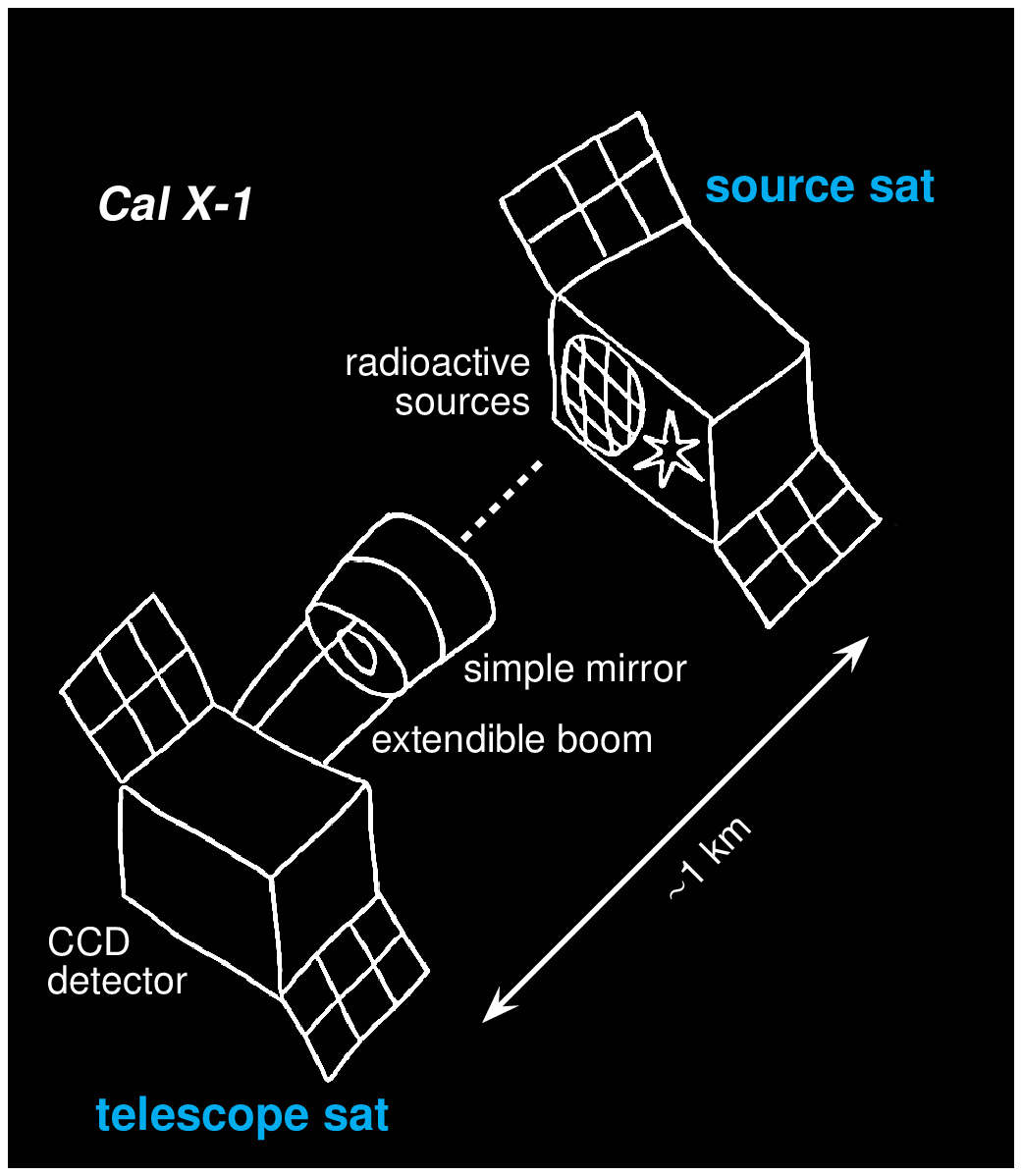}}

\rput[bl]{0}(8.1,0.){%
\includegraphics[angle=0,height=7cm,bb=1 1 446 512,clip]%
{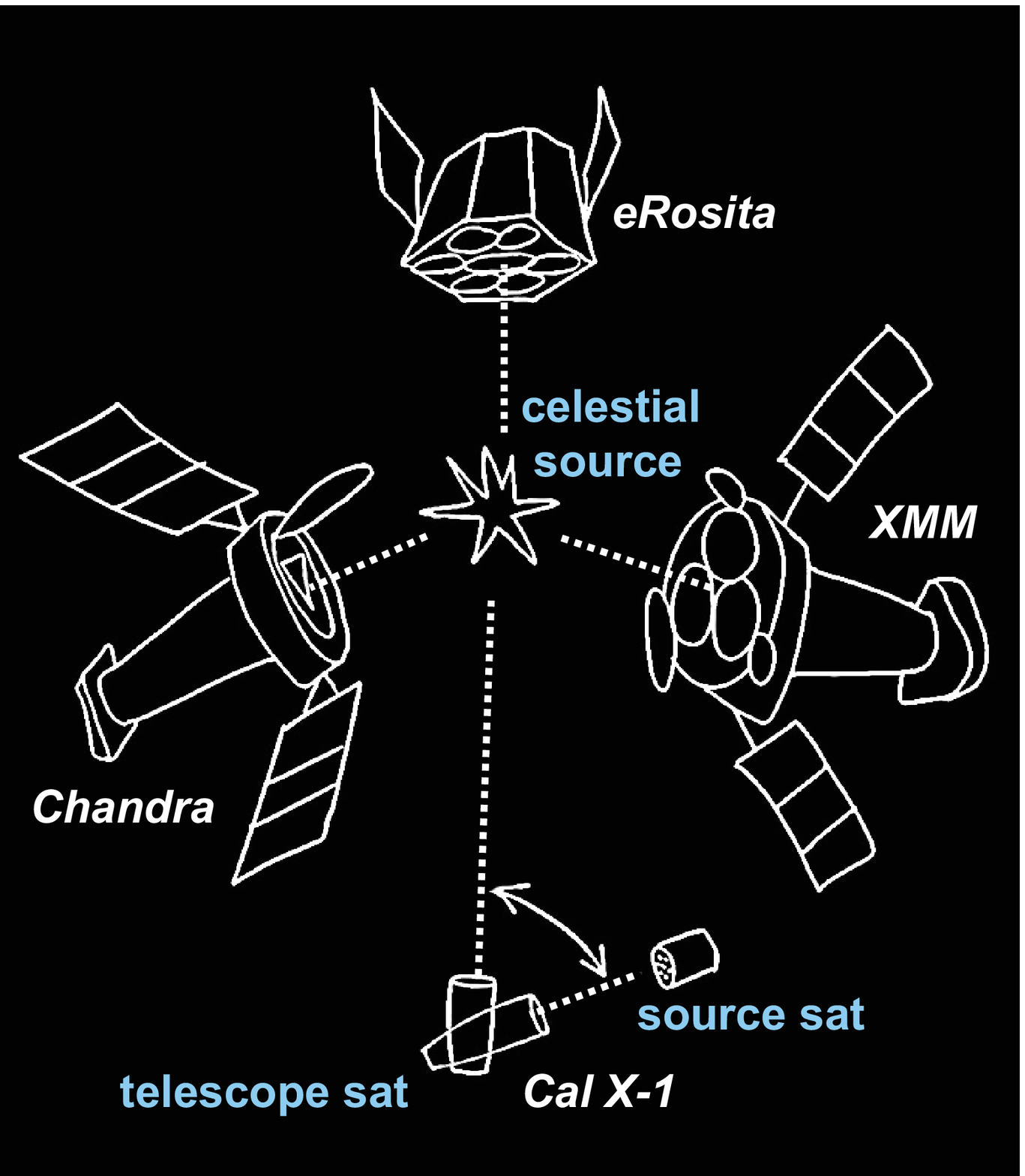}}

%\psgrid(0,0)(18.0,15.5)
\endpspicture

% \caption{\calx\ concept. Calibrtion is transferred from a well-measured
%   radioactive source deployed in space on one of a pair of CubeSats, to a
%   non-variable celestial source (such as supernova remnant), which becomes a
%   standard candle for all X-ray observatories.}
% \label{fig:sketch1}
\end{figure*} 
%%%%%%%%%%%%%%%%%%%%%%%%%%%%%%%%%%%%%%%%%%%%%%%%%%%%%%%%%%%%%%%%%%%%%%%%%%

\clearpage \twocolumn
\setcounter{page}{1}

\section{SUMMARY}

X-ray astronomy is now making observations relevant to fundamental physics
and cosmology. However, the constraining power of many of these measurements
is limited by the instrument calibration uncertainty. Cross-comparison of
the most powerful X-ray observatories in existence, which have also had the
most systematic and thorough ground calibration programs ever, reveal
systematic discrepancies in the measured source fluxes at a $>10$\% level.
Clearly, ground calibration alone cannot provide a better accuracy because
of its inherent limitations and the changes that the observatories
experience on their way to orbit. It is unclear which observatory (if any!)
is correct, because there are no ``standard candles'' in the X-ray sky.

We embark on a task to establish, for the first time, such celestial
standard candles and calibrate their X-ray fluxes to a 1--2\% absolute
accuracy. The absolute calibration will be performed by a {\bs\small
  SmallSat}-sized mission consisting of two satellites flying in formation.
A 6U-sized CubeSat will carry an absolutely calibrated X-ray source, and a
companion 12U CubeSat will carry a small, simple X-ray telescope. The
satellites will have separation large enough to be effectively infinite (for
telescope illumination) and with precisely known distances (to predict the
flux).  Observations of the calibrated source, interleaved with observations
of strategically chosen, non-variable celestial sources, will transfer the
absolute calibration of \calx\ to the celestial sources.

This will allow true end-to-end, in-orbit effective area calibrations of
current and future X-ray observatories, reducing an important systematic
uncertainty in the interpretation of X-ray observations. The new calibration
will apply to vast archives of \chandra, \xmmnewt, \suzaku, \swift, \nustar,
\nicer\ and other X-ray observatories working at 1--10 keV energies, as well
as future data. Thus, at a very modest cost, \calx\ will significantly
enhance the scientific value of billion-dollar missions.

The \calx\ mission requires precision calibration of radioactive sources,
precise relative navigation of two CubeSats, a miniaturized extendable boom,
an efficient but vignetting-free X-ray mirror, and a CubeSat-compatible
X-ray CCD camera.  Each of these technologies enables \calx, and each
technology is attainable building upon existing capabilities.

\section{NEED FOR ABSOLUTE X-RAY FLUXES}

%%%%%%%%%%%%%%%%%%%%%%%%%%%%%%%%%%%%%%%%%%%%%%%%%%%%%%%%%%%%%%%%%%%%%%%%%%
\begin{figure}[b]
\centering
\pspicture(0,0.2)(8,7)

\rput[bl]{0}(0.,0.){%
\includegraphics[angle=0,width=7.5cm]%
{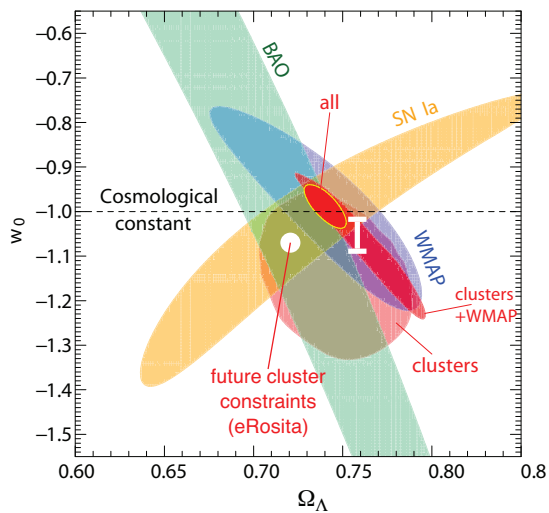}}

%\psgrid(0,9)(18.0,15.5)
\endpspicture

\caption{\small Constraints on Dark Energy density and equation of state
  from different methods.  Current cross comparisons are limited by
  systematic errors in absolute calibration of X-ray flux measurements.
  \calx\ will reduce this systematic below the statistical error of the next
  generation of X-ray surveys.}
\label{fig:cosmol}
\end{figure} 
%%%%%%%%%%%%%%%%%%%%%%%%%%%%%%%%%%%%%%%%%%%%%%%%%%%%%%%%%%%%%%%%%%%%%%%%%%

Many astrophysical measurements of fundamental importance rely on accurate
effective area calibration of X-ray telescopes, such as \chandra\ and
\xmmnewt. A particularly important case is cosmological constraints based on
galaxy clusters. Discovering how the Universe formed and what natural forces
control its evolution is one of the most basic astronomical endeavors. The
cosmological model, which quantifies the various types of matter (baryonic
and dark matter, massive neutrinos, dark energy) that control the geometry
and expansion rate of the Universe, can be constrained in several ways. As
an example, Fig.\ \ref{fig:cosmol} shows relatively recent constraints on
two interesting cosmological parameters, the Dark Energy density,
$\Omega_\Lambda$, and its equation of state parameter, $w_0$, derived from
the Cosmic Microwave Background (CMB, labeled WMAP in the figure), Type 1a
supernovae (SN 1a), Baryonic Acoustic Oscillations (BAO), and clusters of
galaxies. The distribution of masses of clusters --- the most massive
gravitationally bound objects in the Universe with masses $\sim
10^{15}$\msun --- is very sensitive to the cosmological model.  Clusters
probe cosmology in the low-$z$ universe, while the CMB traces the state of
the Universe at its dawn ($z=1000$). The `clusters' constraints in Fig.\ 
\ref{fig:cosmol} are derived from \chandra\ X-ray mass measurements
(Vihklinin et al.\ 2009) and are complementary to other methods.
Combinations of all the different methods could provide the most stringent
constraints (labeled `all') if all measures agree --- or indicate the need
for new physics if disagreements persist. Indeed, the most recent studies
hint at tension between cluster and CMB constraints, and one proposed
explanation is unexpectedly massive neutrinos (Planck Collaboration 2016). It is
thus vitally important to exclude the measurement errors.

Current cluster constraints are still limited by small-sample statistics,
but forthcoming large X-ray emission and Sunyaev-Zeldovich (SZ) microwave
background decrement cluster catalogs (e.g. eRosita, Merloni et al.\ 2012;
SPT-3G, Benson et al.\ 2014) will yield order-of-magnitude tighter
statistical constraints (the expected eRosita constraint is shown in Fig.\ 
\ref{fig:cosmol}). However, the white error bar in Fig.\ \ref{fig:cosmol}
shows the effect of a 10\% systematic error on cluster masses. A bias of
this magnitude will render those more accurate results systematics-limited.
This problem is widely recognized by cosmologists working at different
wavelengths; the Planck Collaboration (2016) noted that ``Improving the
precision of cluster mass calibration from the current 10\% level to 1\%
would ... provide a stringent test of the base $\Lambda$ Cold Dark Matter
model.''  This is a recognition that ``while the X-ray astronomy community
has achieved extraordinary advancements in the accuracy of X-ray
measurements, it seems to hit a ceiling as far as the precision of some of
them is concerned'' (M.\ Guainazzi, IACHEC, Section 12; see also {\bs\small
  Madsen et al.\ Astro2020 White Paper on IACHEC}).

%%%%%%%%%%%%%%%%%%%%%%%%%%%%%%%%%%%%%%%%%%%%%%%%%%%%%%%%%%%%%%%%%%%%%%%%%%
\begin{figure}[t]
\centering
\pspicture(0,0.5)(8,18.5)

\rput[bl]{0}(0.2,0.){%
\includegraphics[angle=0,width=7.35cm]%
{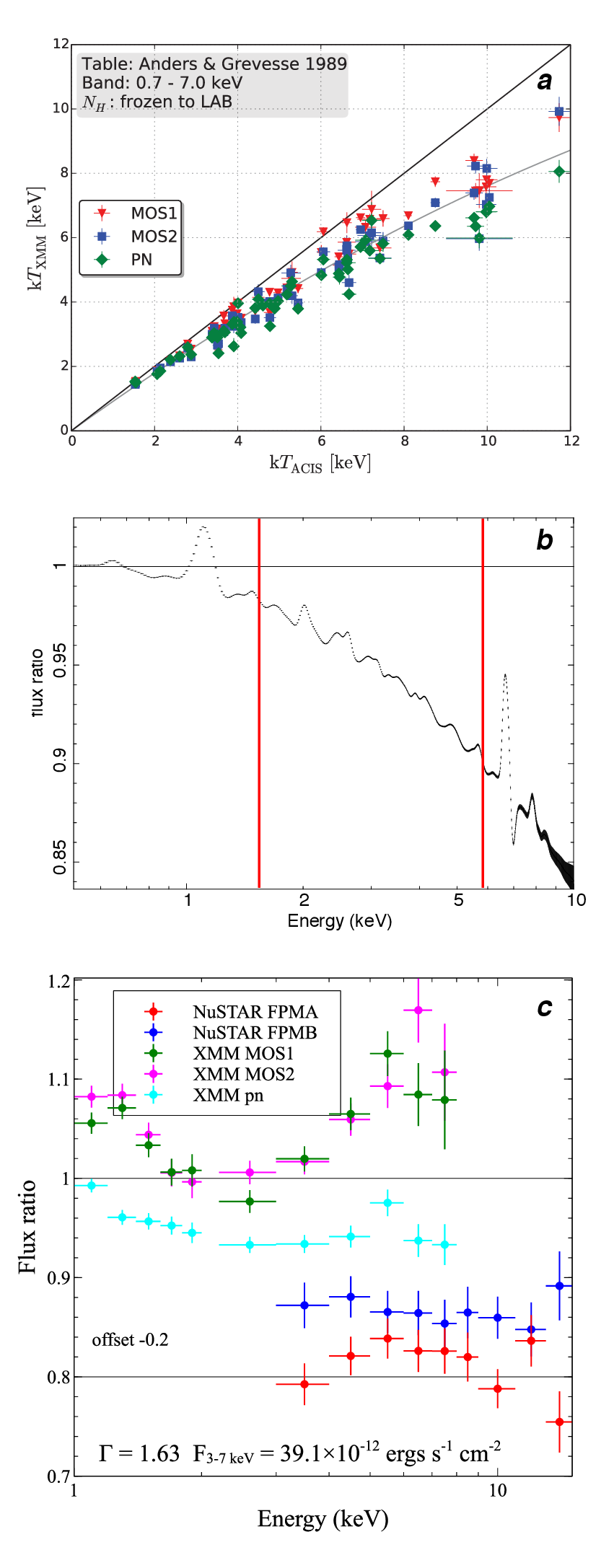}}

%\psgrid(0,9)(18.0,15.5)
\endpspicture

\caption{\small (a) Cluster temperatures from \chandra\ and XMM
  (Shellenberger et al.\ 2015) show significant discrepancies. (b) The ratio
  of model spectra for T=5.0 keV and 5.5 keV.  A 10\% difference in
  temperature corresponds to $\sim10$\% different ratio of fluxes around 1
  keV and 5 keV. (c) The ratio of spectra of 3C273 to a common model for
  \xmm\ and \nustar\ instruments (Madsen et al.\ 2017). There are large
  differences in derived fluxes for the same source observed at the same
  time.}
\label{fig:temps}
\end{figure} 
%%%%%%%%%%%%%%%%%%%%%%%%%%%%%%%%%%%%%%%%%%%%%%%%%%%%%%%%%%%%%%%%%%%%%%%%%%

A cluster total mass that is used in the cosmological tests, $M_{\rm tot}$,
can be determined by several methods affected by different systematics. It
can be estimated most easily and reliably from the hydrostatic equilibrium
method and X-ray derived gas temperature, $T$. Cluster temperatures
currently derived by different X-ray instruments are systematically
discrepant. Figure \ref{fig:temps} shows the comparison of the temperatures
from the \chandra\ ACIS and \xmm\ EPIC cameras (both MOS and PN) for the
same clusters (Shellenberger et al.\ 2015).  For the hottest (and most
cosmologically constraining) clusters, the difference between the \xmm\ and
\chandra\ derived temperatures reaches 20\%.  As seen from Fig.\ 
\ref{fig:temps}b, which shows the ratio of model spectra for thermal plasmas
with $T=5$ keV and 5.5 keV folded through the mirror and detector response,
the temperature difference stems from the difference in spectral slope in
the 1--10 keV band. For temperatures in this range, a 10\% error in the
ratio of fluxes around $E=1$ keV and 5 keV would result in a 10\%
temperature error. Figure \ref{fig:temps}c shows the spectra of \xmm\ and
\nustar\ detectors for simultaneous observations of 3C273 (Madsen et al.\ 
2017).  Indeed, there are 10--20\% energy-dependent differences between the
instruments; differences with \chandra\ are of similar magnitude.  These
discrepancies are due to errors in the energy-dependent system efficiencies
of at least one (or, more likely, all) of the large observatories.  These
discrepancies limit the cosmological studies using X-ray derived galaxy
cluster masses. Qualitatively, $M_{\rm tot}\propto T$, so a 10\% error in
$T$\/ results in a 10\% mass error.

Other fundamental measurements, which span the range of observational X-ray
astronomy from black holes and neutron stars to clusters of galaxies and the
cosmic X-ray background, also rely on absolute X-ray flux calibration. We
mention some of them briefly.

The classic geometric Hubble constant test derives distances to clusters of
galaxies by comparing their X-ray temperatures and fluxes to SZ microwave
decrements (Silk \& White 1978). The distance to a cluster, $d_a \propto
y^2/(f_X T^2)$, where $y$\/ is the SZ signal and $f_X$ is the X-ray flux. A
large uncertainty in these measurements stems from the X-ray calibration for
$f_X$ and $T$. (The uncertainty in $y$\/ is beyond our scope.) If one
chooses to believe that cosmology will be solved by other methods (Fig.\ 
\ref{fig:cosmol}), the above distance measurement enables a measurement of
the Helium abundance in clusters (Markevitch 2007).  Cluster cosmological
tests rely on the fundamental, if unstated, assumption that cluster He
abundance is at the universal average, while certain very consequential
physical processes could modify it.

Determining the equation of state of the ultra-dense matter of the neutron
stars requires the neutron star's radius and mass. \nicer\ will provide one
measure via detailed fitting of the shape and energy dependence of the phase
folded light curve; a complementary measure comes from the absolute X-ray
flux of Eddington luminosity of Type I X-ray bursts.  A $<3$\% absolute flux
accuracy is required to start distinguishing between the interesting
equations of state (Ozel et al.\ 2016).

Measurements of black hole spin are made by fitting thermal continuum
spectra to the accretion disk spectra of the soft state and by fitting the
shape of the relativistically broadened Fe line in the reflection spectra of
the hard state.  These methods rely on the strong dependence of the
innermost stable radius of the accretion disk on black hole spin.  For the
continuum method, X-ray flux is proportional to the solid angle filled with
emitting material, which depends on both inclination and inner radius.
Uncertainties in the absolute X-ray flux thus contribute directly to the
systematic errors in derived spin (Steiner et al.\ 2014).

At the other end of the observational spectrum, the intensity of the Cosmic
X-ray Background may provide a measure of the history of gravitational
accretion integrated over cosmic time (Boldt \& Leiter 1995).

%%%%%%%%%%%%%%%%%%%%%%%%%%%%%%%%%%%%%%%%%%%%%%%%%%%%%%%%%%%%%%%%%%%%%%%%%%
\begin{figure*}[t]
\centering
\pspicture(0,0)(15,6)

\rput[bl]{0}(-0.6,0.){%
\includegraphics[angle=0,height=5.8cm,bb=51 460 389 643,clip]%
{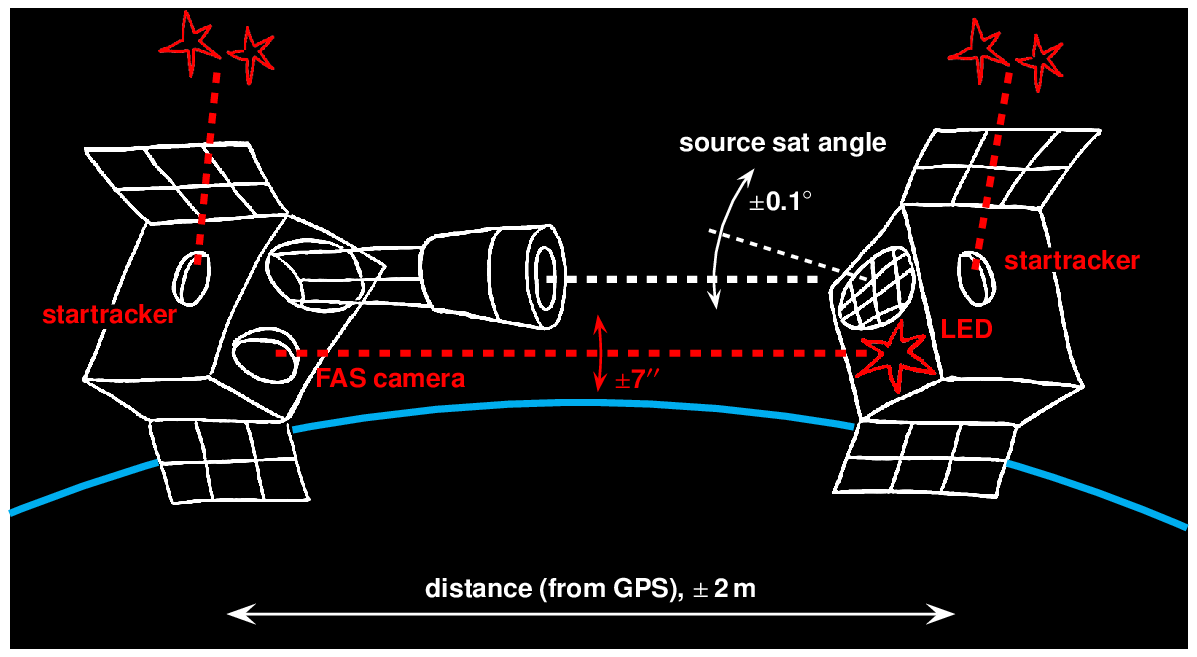}}

\rput[bl]{0}(10.5,0.){%
\includegraphics[angle=0,height=5.8cm,bb=146 475 295 644,clip]%
{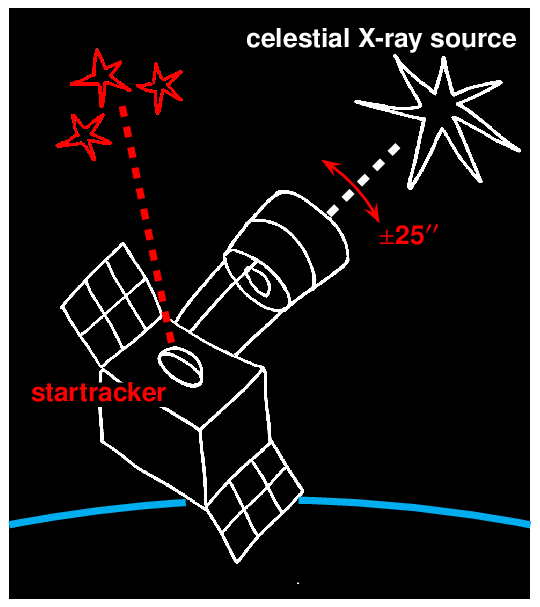}}

%\psgrid(0,9)(18.0,15.5)
\endpspicture

\caption{\small \calx\ (left) interleaves observations of an absolutely
  calibrated source with observations of the celestial X-ray standard
  candles. The standard candles, for example, supernova remnants G21.5--0.9
  and N132D, are known to be time invariant, small in angular extent,
  sharply bounded, and relatively bright. The major unknown property of
  these standard candles is their absolute brightness, which \calx\ 
  observations will determine.}
\label{fig:sketch2}
\end{figure*} 
%%%%%%%%%%%%%%%%%%%%%%%%%%%%%%%%%%%%%%%%%%%%%%%%%%%%%%%%%%%%%%%%%%%%%%%%%%

\section{WHAT'S WRONG WITH GROUND CALIBRATION?}

\chandra\ and \xmmnewt\ were both calibrated extensively prior to launch, with the
goal of a $\sim1$\% accuracy. However, a $>10$\% discrepancy remains between
the observatories even after attempts to reconcile the instruments over the
last 15 years.  Among the challenges of ground calibration is that the
illumination pattern of a large-diameter X-ray mirror on the ground differs
from the illumination by a celestial source at infinity.  Celestial cross
calibrations indicate discrepancies but cannot by themselves determine which
calibration is closer to the truth. The on-board radioactive sources
employed by current missions to monitor time-dependent effects in the
detectors and filters cannot monitor time dependent changes to the mirror.
An in-orbit, absolute, calibration standard is needed to provide both a
calibration and a monitor of time-dependent system changes.

\section{CAL X-1 CONCEPT}

Our \calx\ mission concept aims to provide two (or more) X-ray standard
candles in the sky that could be used by current and future soft X-ray
observatories, at a very modest cost. It will establish the standard candles
at 5.9 keV, the Mn K$\alpha$ line energy produced by the radioactive
$^{55}$Fe source, and at 1.5 keV (Al K$\alpha$ line), providing calibrations
above and below the atomic edge features present in all X-ray mirrors near 2
keV. In addition to providing the absolute fluxes, these are pivotal
energies to pinpoint the spectral slope that determines the gas temperatures
(Fig.\ \ref{fig:temps}b). Our goal is 2\% flux accuracy at each energy,
which will be a substantial leap forward from the current state of the art.

\chandra, \xmmnewt\ and other big observatories have been meticulous in
calibrating the relative time dependence of their effective areas over their
lifetimes by periodically observing constant celestial sources, including
our primary targets. Thus a true standard candle will allow the calibration
of their entire data archives. Even a \calx\ measurement in $\sim 2023$ can
be applied to today's missions.

As X-ray telescopes become more powerful, the demands on calibration
increase. The in-orbit effective area calibration pioneered by \calx\ will
be relevant to supporting large missions currently being built (\athena,
\xrism) or under study (\lynx, \axis).

\section{TECHNICAL APPROACH}

We propose a CubeSat mission as a low-cost ($<$\$20M) solution to improve
the calibration of Observatory-class missions (>\$1B).  \calx\ consists of
two CubeSats flying in formation, which alternately perform self-calibration
operations and observations of key celestial sources.

One \calx\ satellite (SrcSat) carries absolutely calibrated X-ray sources,
including a bright radioactive source for 5.9 keV and a modulated X-ray
source for 1.5 keV. A companion satellite (TelSat), held at a distance of
1--2~km, carries a small X-ray telescope. The telescope is alternately
pointed to SrcSat and a celestial source (Fig.\ \ref{fig:sketch2}). The
primary objective of \calx\ is to observe two celestial sources that are
already widely observed for cross calibration purposes, thus enabling the
transfer of the \calx\ calibration to other observatories.

The mission is {\bs\small designed from the ground up to cancel out or
  minimize any effects of \calx\ own systematics} on the celestial source
calibration. For example, the X-ray mirror uses a special design that
provides a large vignetting-free area around the optical axis (Fig.\ 
\ref{fig:vign_boom}), greatly relaxing the pointing accuracy requirements.

The two satellites do not need to communicate with each other. The precision
distance required for determining the absolute SrcSat flux is provided by
GPS; precision pointing to the SrcSat is derived from a beacon on SrcSat and
a camera on TelSat (LED beacon and FAS camera in Fig.\ \ref{fig:sketch2}).
During celestial source observations, startrackers that come with the
commercially procured CubeSat buses provide the required accuracy of
attitude control. TelSat has a propulsion system to maintain coarse control
of satellite separation.

%%%%%%%%%%%%%%%%%%%%%%%%%%%%%%%%%%%%%%%%%%%%%%%%%%%%%%%%%%%%%%%%%%%%%%%%%%
\begin{figure}[b]
\centering
\pspicture(0,0.2)(15,11.3)

\rput[bl]{0}(0,0.){%
\includegraphics[angle=0,width=7.5cm,bb=18 175 549 683,clip]%
{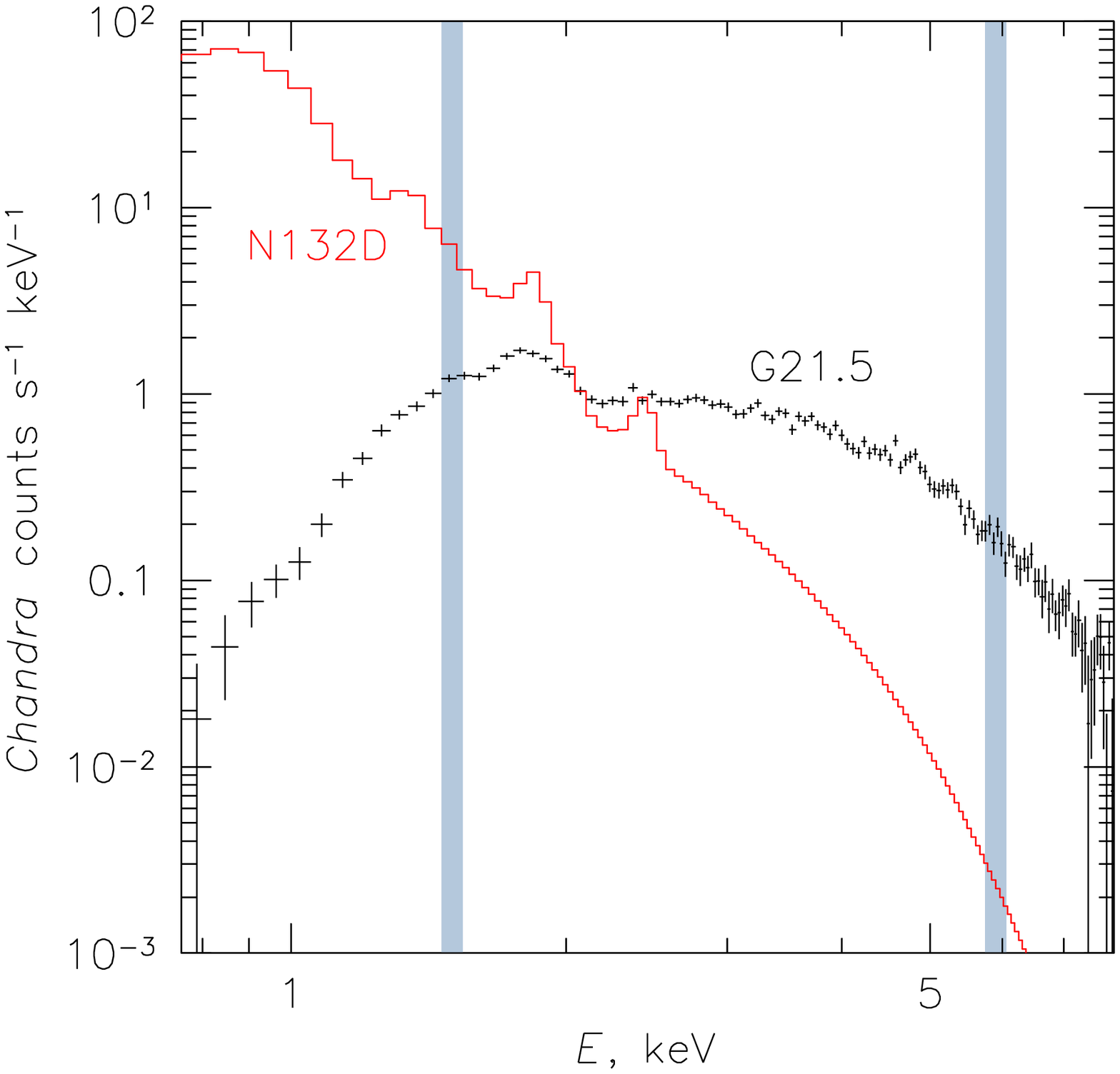}}

\rput[bl]{0}(0.2,7.6){%
\includegraphics[angle=0,height=3.4cm,bb=10 3 190 170,clip]%
{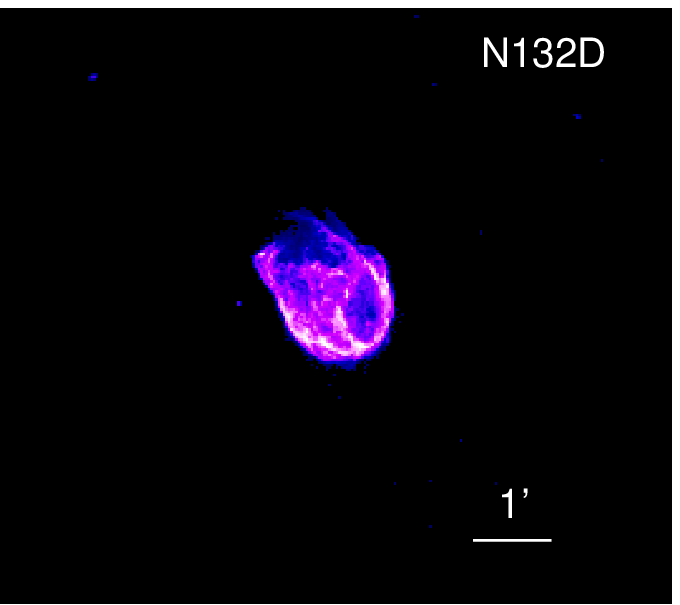}}

\rput[bl]{0}(4.1,7.6){%
\includegraphics[angle=0,height=3.4cm,bb=10 3 190 170,clip]%
{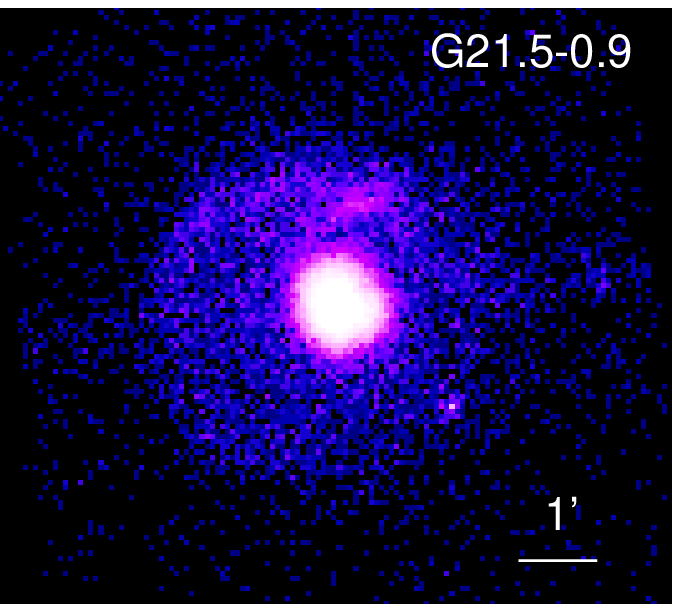}}

%\psgrid(0,0)(8,11)
\endpspicture

\caption{\small \chandra\ spectra and images of the future X-ray standard
  candles. The \calx\ calibration energies are indicated by the blue
  vertical bars.}
\label{fig:snr}
\end{figure} 
%%%%%%%%%%%%%%%%%%%%%%%%%%%%%%%%%%%%%%%%%%%%%%%%%%%%%%%%%%%%%%%%%%%%%%%%%%

\subsection{X-ray Standard Candles}

The International Astronomical Consortium for High Energy Calibration
(IACHEC, Madsen et al.\ 2019) has identified the supernova remnants (SNR)
G21.5--0.9 and N132D as good, time stationary, X-ray candles for cross
calibration.  While galaxy clusters are also constant, they are fainter and
lack well-defined boundaries, making the finite telescope angular resolution
the source of an additional uncertainty. By contrast, these SNRs are limited
in angular extent, sharply bounded, and have suitable brightness for both
large and small telescopes. Their \chandra\ images and spectra are shown in
Fig.\ \ref{fig:snr}. These two SNRs have been used for in-flight effective
area monitoring and cross-calibration of most major Observatories working in
the 1--10 keV range.  They were among the 6 verification targets observed by
\hitomi, and will undoubtedly be used by future X-ray instruments.

We will use N132D for the 1.5 keV calibration and G21.5, which is brighter
at higher energies (Fig.\ \ref{fig:snr}), at 5.9 keV. G21.5 provides an
additional measurement at 1.5 keV, albeit with lower statistical accuracy.
N132D has a radius of $\sim 1'$; G21.9 has an outer radius of $\sim 3'$, but
85\% of the flux is contained in the inner core, and the remnant has an
effective radius of $\sim 1'$.  These sources are the subject of IACHEC
sponsored cross-calibrations (Tsujimoto et al.\ 2011; Pollock \& Guainazzi
2014).

\calx\ will observe each source for $10^6$\,s to achieve the required
statistical accuracy. The exposures will be interleaved with SrcSat
exposures. We will measure the source fluxes in narrow intervals
(commensurate with the resolution of the CCD detectors of \calx\ and other
observatories) around the two pivotal energies.  Technically, the spectral
models (developed by IACHEC) will be adjusted to describe both our G21.5 and
N132D observations and the SrcSat monochromatic lines, thereby {\bs\small
  canceling out} the calibration of \calx\ own instruments.  G21.5 has a
featureless absorbed power-law spectrum (\hitomi\ Collaboration 2018), while
N132D exhibits multiple emission lines (Plucinsky et al.\ 2018), which will
not be a concern given the well-developed model for those lines and the
design of our measurement. If the mission lives longer than planned, more
celestial sources can be observed.

\subsection{X-ray sources}

SrcSat carries precisely calibrated $^{55}$Fe sources (producing the 5.9 keV
Mn K$\alpha$ line) and a Modulated X-ray Source (MXS) (Gendreau et al.\ 
2012) with lines at 5.9 keV and 1.5 keV, generated by an anode constructed
of Mn and Al. It is possible to use Al fluorescence (1.5 keV) induced
directly by $^{55}$Fe illumination, but the resulting 1.5 keV flux would be
too low, hence our choice of MXS, which can provide a much brighter 1.5 keV
line.

To provide the required line brightness for TelSat, the $^{55}$Fe sources
will have the total activity of 0.16 Ci, consisting of an array of 12
standard industrial sources in vacuum-qualified steel capsules with a Be
window. The absolute intensity of each of these sources will be calibrated
on the ground at National Institute of Standards and Technology (NIST) using
their radioactive standard. NIST routinely calibrates radionuclidic sources
to uncertainties on the order of 1\%. Thereafter, the time dependence of the
source flux is a simple exponential decay curve with a precisely known
half-life of 2.7 yr, independent of their environment.

The ratio of the two MXS line intensities is stable and well-known --- from
a calibration at NIST --- as a function of measurable operating conditions
such as voltage and temperature.  MXS will be turned on and off, while the
radioactive 5.9 keV source is always on. Under the assumption of the known
ratio of MXS line intensities, the intensity of the 1.5 keV line is
calibrated by comparing the total 5.9 keV intensities when the MXS is on
(MXS + $^{55}$Fe) and off ($^{55}$Fe only). The MXS will be built at Goddard
using commercial UHV components. 

The ratio of 1.5 keV to 5.9 keV lines could be affected by organic
contamination build-up on the MXS exit window after its calibration. The
storage of the source in a dry N$_2$ environment will limit such
contamination to well below the \calx\ requirement ($<0.5\mu$m of organic
contamination build-up, corresponding to a transmission at 1.5 keV of
$>0.993$).

%%%%%%%%%%%%%%%%%%%%%%%%%%%%%%%%%%%%%%%%%%%%%%%%%%%%%%%%%%%%%%%%%%%%%%%%%%
\begin{figure*}
\centering
\pspicture(0,0)(15,5.8)

\rput[bl]{0}(-0.5,0.){%
\includegraphics[angle=0,height=5.5cm,bb=154 73 490 581,clip]%
{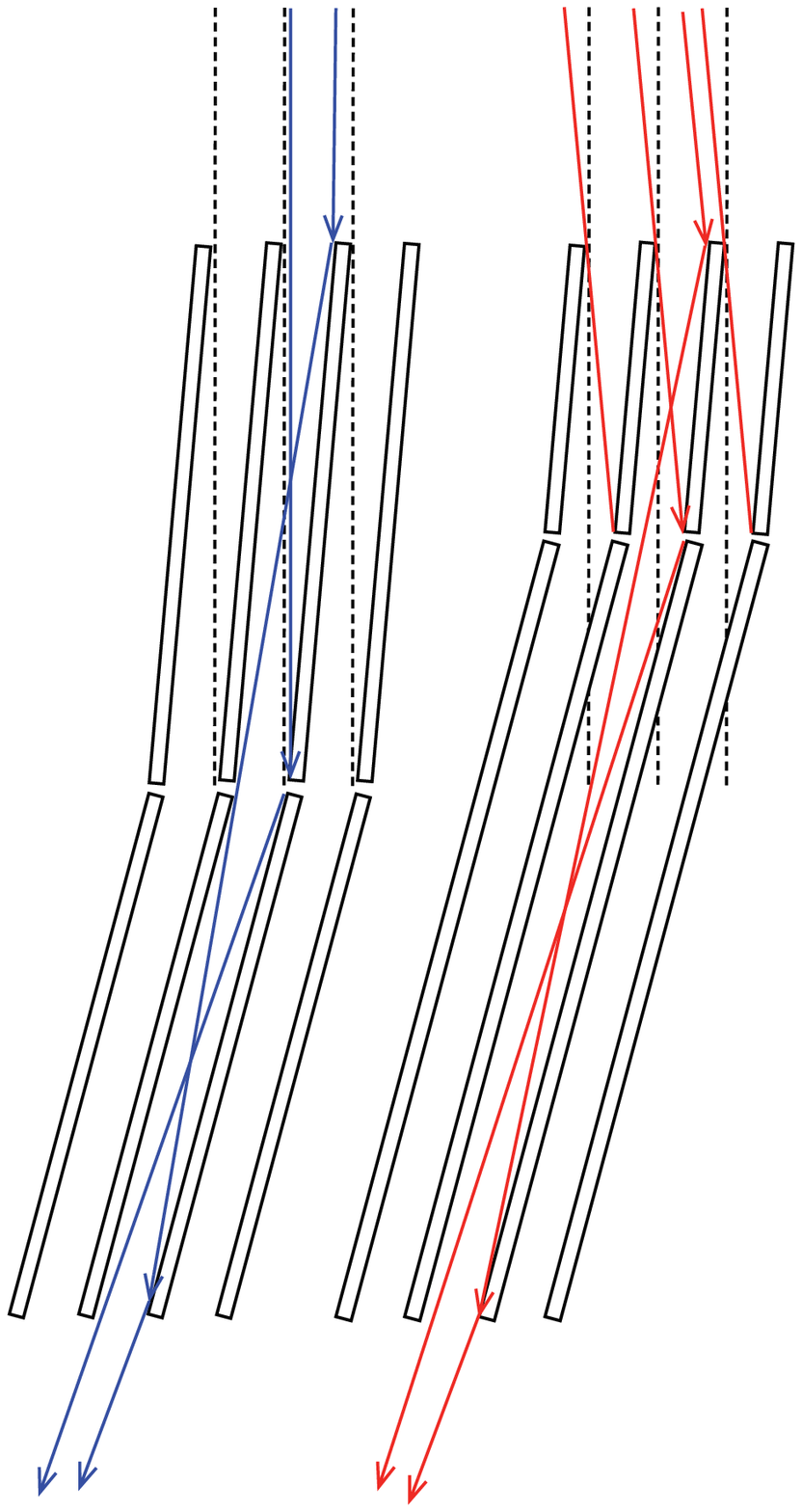}}

\rput[bl]{0}(3.6,0.){%
\includegraphics[angle=0,height=5.5cm,bb=1 3 162 123,clip]%
{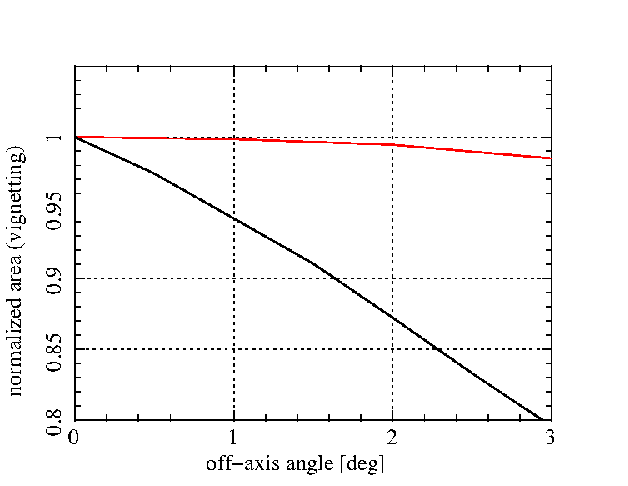}}

\rput[bl]{0}(11.8,0.7){%
\includegraphics[angle=0,height=4.7cm,bb=1 1 114 136,clip]%
{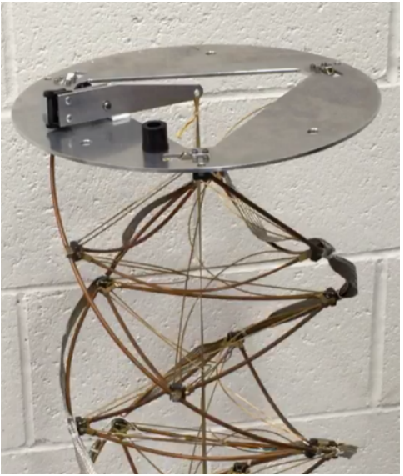}}

\rput[bl]{0}(0,5.0){\fsz\bsi a}
\rput[bl]{0}(4.7,5.0){\fsz\bsi b}
\rput[bl]{0}(12,5.0){\fsz\bsi c}

%\psgrid(0,0)(18.0,15.5)
\endpspicture

\caption{\small The modification to primary/secondary X-ray mirror length
  ({\em a}; left is traditional design and right is modified design), which
  enables a vignetting-free central field of view (red curve in {\em b}) at
  the expense of some on-axis effective area. Black curve in ({\em b}) is
  for standard design. ({\em c}) A partially deployed coilable boom of the
  type that will be used to deploy the \calx\ mirror, made by Orbital ATK.}
\label{fig:vign_boom}
\end{figure*} 
%%%%%%%%%%%%%%%%%%%%%%%%%%%%%%%%%%%%%%%%%%%%%%%%%%%%%%%%%%%%%%%%%%%%%%%%%%

\subsection{Telescope components}

\calx\ carries an X-ray CCD camera at the focus of a small grazing incidence
mirror supported by a one-time extendable boom. The mirror is designed from
the ground up to minimize the systematics of our measurement and allow for
the finite source size, pointing accuracy, and boom stability by providing a
large vignetting-free region ($>99$\% for $r<2'$, Fig.\ 
\ref{fig:vign_boom}). True X-ray imaging with $\sim1'$ PSF and the
$15'\times11'$ detector field of view ensure that we minimize the
uncertainty from the PSF-scattered flux and accurately evaluate the (low)
detector background.

{\bi Mirror.}~~ The X-ray mirror is based on the well-established designs of
\hitomi\ and \nicer\ (both built at Goddard). For \calx, we sacrifice some
effective area in order to reduce vignetting. This is achieved by shortening
the primary and lengthening the secondary reflectors and having larger
spacing between adjacent shells.  Fig.\ \ref{fig:vign_boom}{\em a}\/ shows
the modification to the lengths and the resulting vignetting curve.  A
mirror of this design with 18cm diameter and 11cm height (which can be
packed into a CubeSat pre-deployment) has an effective area of 54 cm$^2$ at
1.5 keV and 43 cm$^2$ at 6 keV.

{\bi Coilable boom.}~~ Orbital ATK manufactures deployable structures and
coilable booms that are rigid and accurate enough to be part of astronomical
telescope systems (McEachen 2011, 2013; Fig.\ \ref{fig:vign_boom}c). They
will supply a 2m boom that meets \calx\ requirements for stability. Our
concept of operations cancels out a small uncertainty of the initial
placement as deployed in orbit.  Pre-deployment, the boom is packed down to
2cm in height inside the CubeSat.

{\bi Detector.}~ CCD is the workhorse of X-ray astronomy, providing a good
compromise between spatial and spectral resolution.  XCAM Ltd.\ has
developed and flown optical imaging detectors on two successful CubeSats
(Harris et al.\ 2011; AlSat 2016), and have experience with X-ray CCD camera
systems for other missions. One of those CubeSats also contained a
thermoelectric cooler to maintain the CCD temperature as part of the same
experiment. For \calx, we will use XCAM's off-the-shelf $385\times 288$
pixel CCD, which will provide a $15'\times 11'$ field of view.

A proton shield and collimator with a small aperture will protect the CCD
from the proton background and from stray light effects. Based on \chandra\ 
and \swift\ experience, we estimate the CCD background to be much lower than
the source brightness at both energies, with small contributions to the
error budget, even with a conservatively high estimate of a possible
background Al fluorescent line (Table 1).

\calx\ is relatively insensitive to CCD efficiency or even contamination
build up on the CCD.  Both are canceled out by the observations which
alternate between the calibration source and celestial source.

\subsection{Relative Navigation System}

The relative navigation system, employed when TelSat is pointed at SrcSat,
is not yet available from the commercial CubeSat community, so it will be
built at Goddard. It consists of a beacon on SrcSat, a sensor on TelSat, and
an interface to the spacecraft attitude control system (ACS). SrcSat will
have an array of LEDs which is unresolved at 1 km. The brightness is the
equivalent of a magnitude 3 star at a distance of 2 km. The beacon will be
pulsed at 1 Hz with a duty cycle of <10\%. The small duty cycle allows X-ray
data collection when the beacon is off, eliminating the possibility of
optical background from the beacon itself.

TelSat contains a Fine Alignment Sensor (FAS) co-aligned with the X-ray
telescope. To exclude parallax, the distance between the X-ray telescope and
FAS is the same as that between the X-ray source and the LEDs on SrcSat.
The sensor has a field of view of 9\deg\ and the ability to centroid the LED
beacon signal to $<7''$. At the SmallSat scale, FAS is similar to the
navigation system flown on the Swedish PRISMA mission. At the CubeSat scale,
the FAS is similar to the optical alignment system on CANYVAL-X (Park et
al.\ 2016).

In operation, SrcSat will point the X-ray sources and beacon anti-parallel
with the direction of motion.  Accuracy of better than 1\deg\ is required
and easily obtained.  TelSat will acquire the beacon signal and shift to an
ACS control mode, which points TelSat towards SrcSat.

The \calx\ error budget that includes all systematics (many of which are not
discussed in this paper) is shown in Table 1.  We will be able to achieve
the goal of 2\% flux accuracy.

%%%%%%%%%%%%%%%%%%%%%%%%%%%%%%%%%%%%%%%%%%%%%%%%%%%%%%%%%%%%%%%%%%%%%%%%%%
\begin{figure*}
\centering
\pspicture(0,0)(15,8.4)

\rput[bl]{0}(-0.8,0.){%
\includegraphics[angle=0,height=8.3cm]%
{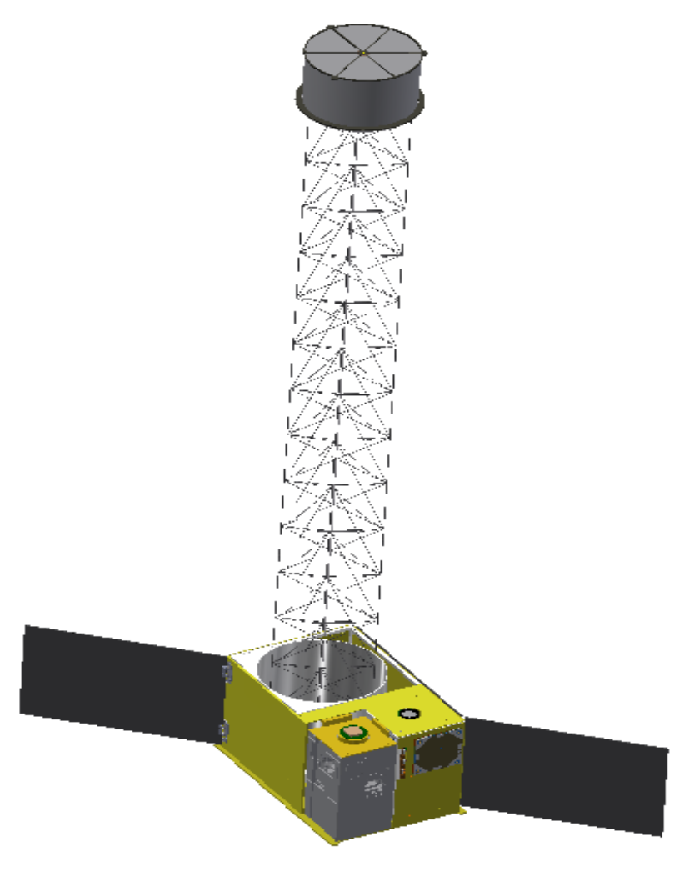}}

\rput[bl]{0}(6.6,6.0){%
\includegraphics[angle=0,height=2.0cm]%
{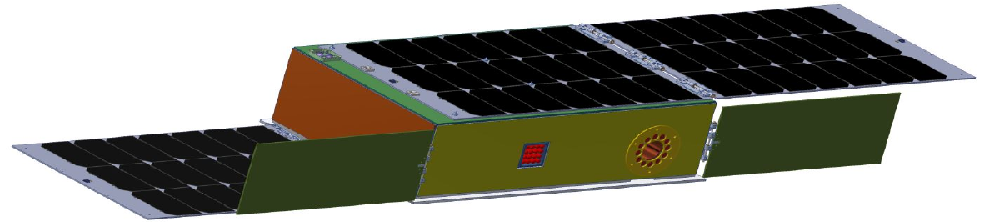}}

\rput[bl]{0}(6.6,0.){%
\includegraphics[angle=0,height=5.0cm]%
{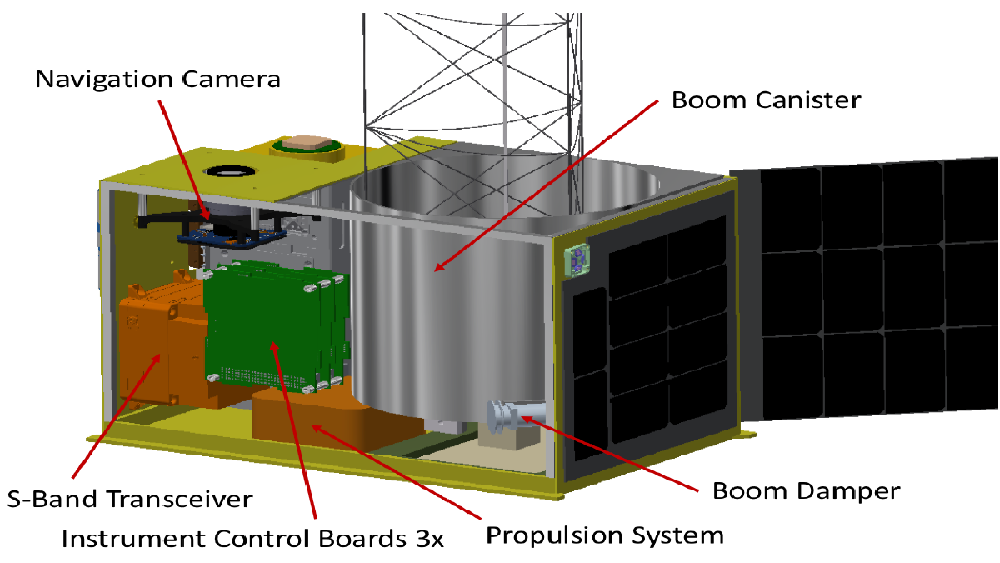}}

\rput[bl]{0}(1.5,7.5){\fsz\bsi a}
\rput[bl]{0}(7.0,7.5){\fsz\bsi b}
\rput[bl]{0}(15,4.8){\fsz\bsi c}

%\psgrid(0,0)(18.0,15.5)
\endpspicture

\caption{\small \calx\ mission design by NASA Wallops satellite lab. ({\em a,c})
  TelSat carries an X-ray mirror on a deployable boom, CCD detector, FAS
  sensor, an S-band patch antenna, and a micro-propulsion unit, in a 12U
  CubeSat bus from Blue Canon. ({\em b}) SrcSat includes radioactive
  sources, the modulated X-ray source, and a cluster of LEDs for fine
  alignment with the telescope, all on the same spacecraft face of a 6U
  CubeSat. Both satellites have deployable solar panels.}
\label{fig:cad}
\end{figure*}
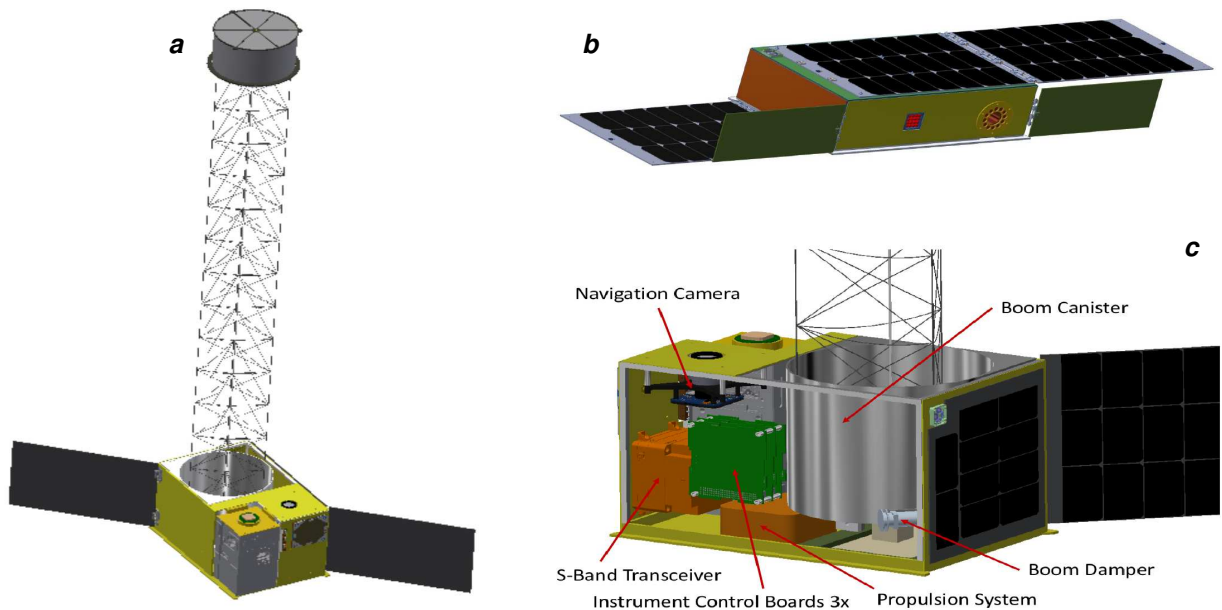 
%%%%%%%%%%%%%%%%%%%%%%%%%%%%%%%%%%%%%%%%%%%%%%%%%%%%%%%%%%%%%%%%%%%%%%%%%%

%%%%%%%%%%%%%%%%%%%%%%%%%%%%%%%%%%%%%%%%%%%%%%%%%%%%%%%%%%%%%%%%%%%%%%%%%%
\begin{table}[t]
\small
\noindent
\hspace*{-1mm}
\renewcommand{\tabcolsep}{2mm}
\renewcommand{\arraystretch}{1.11}
\begin{tabular}{p{4.5cm}cc}
\multicolumn{3}{c}{\normalsize Table 1. {\em Cal X-1}\/ error budget\phantom{\raisebox{-2mm}{\rule{0cm}{8mm}}}} \\
\hline
\hline
 & \multicolumn{2}{c}{Flux uncertainty$^1$} \\
\cline{2-3} &                                    1.5 keV & 6 keV \\
\hline
\multicolumn{3}{c}{\sc Science Requirement \phantom{\huge I}} \\
Delivered accuracy of \\ ~~~~celestial source flux \dotfill & 2.0\%     & 2.0\%   \\
\hline
\multicolumn{3}{c}{\sc Expected Error Budget ($L=1$\,km) \phantom{\huge I}} \\
\multicolumn{3}{c}{\phantom{\huge I}Systematic uncertainties:$^1$\phantom{\huge I}}\\
{\em Calibration source:}\phantom{\Huge g} & & \\
Distance $L$ (from GPS) \dotfill 2m &              0.4\%    & 0.4\% \\
Absolute source calibration \dotfill &             1.5\%    & 1.0\% \\
Source sat orientation \dotfill 0.1\deg &          0.1\%    & 0.1\% \\
\multicolumn{2}{l}{Vignetting (off-axis angle):} & \\
~~~~Source radius$^2$ \dotfill          10\as & & \\
~~~~Mirror radius$^2$ \dotfill          20\as & & \\
~~~~FAS pointing accuracy \dotfill       7\as & & \\
~~~~Boom tilt stability \dotfill        13\as & & \\
~~~~Total angle uncertainty \dotfill    30\as & 0.4\% & 0.4\% \\ 
Background uncertainty  \dotfill              & 0.5\% & 0.1\% \\
Contamination uncertainty  \dotfill           & 0.7\% & 0     \\
Total cal.\ source systematic \dotfill &        1.8\% & 1.2\% \\         
{\em Celestial source:}\phantom{\Huge I} & & \\
Vignetting contributions: & & \\
~~~~ACS pointing accuracy \dotfill     25\as & & \\
~~~~Boom tilt stability \dotfill       13\as & & \\
~~~~Source radius$^{3,2}$ \dotfill     60\as & & \\
~~~~Total angle uncertainty \dotfill   70\as & 0.5\% & 0.5\%\\ 
Background uncertainty  \dotfill             & 0.3\% & 0.2\% \\
Total cel.\ source systematic \dotfill &       0.6\% & 0.55\% \\         
\hline
\multicolumn{3}{c}{\phantom{\huge I}Statistical uncertainties:$^1$\phantom{\huge I}}\\
Calibration source$^4$, 1\,Ms \dotfill &    0.3\%  & 0.3\% \\
Celestial sources, 1\,Ms each \dotfill &    0.25\% & 0.8\% \\
\hline
\multicolumn{3}{c}{\phantom{\huge I}Systematic + statistical:$^1$\phantom{\huge I}}\\
Calibration source\dotfill &                            1.8\% & 1.2\% \\
Celestial source\dotfill &                              0.7\% & 1.0\% \\
Delivered accuracy (celestial \\ ~~~~+ calibration sources)
                                            \dotfill  & 1.9\% & 1.6\% \\
\hline
\end{tabular}

\vspace*{3mm}
\begin{minipage}{7.9cm}
\small
$^1$ Uncertainties are $1\sigma$. 
$^2$ Conservatively using full size as $1\sigma$.
$^3$ Full radius for N132D; effective radius for G21.5 at 6 keV, accounting
for bright core and faint ring. 
$^4$ Assuming 0.16 Ci at time of measurement.
\end{minipage}
\end{table}
%%%%%%%%%%%%%%%%%%%%%%%%%%%%%%%%%%%%%%%%%%%%%%%%%%%%%%%%%%%%%%%%%%%%%%%%%%

\section{MISSION DESIGN}

A detailed design study was performed at NASA Wallops Mission Planning Lab
(MPL), which specializes on CubeSats. It showed that all mechanical,
thermal, power, data rate, communication, and navigation requirements can be
met using mostly off-the-shelf components commercially available for the
CubeSat community. The layout of the two satellites is shown in Fig.\ 
\ref{fig:cad}.  The two satellite buses, 12U and 6U CubeSats, that include
the ACS reaction wheels and startrackers, are offered off-the-shelf by Blue
Canyon Technology.  The satellites will charge their batteries during
daytime and observe at night, which provides substantial power margin. The
CCD will be maintained at --40{\deg}C using a thermoelectric cooler.

TelSat will have a compressed gas micro-propulsion system. The orbit is
selected so that, given the two satellites' ballistic coefficients, a
separation of 1--2 km can be maintained with 1 propulsion maneuver per week.

The \calx\ orbital requirements are given in Table 2. The altitude is driven
primarily by the formation requirements and the orbit decay. The inclination
is driven by the S-band communication and the desire to use NASA Near-Earth
Network. The length of the mission comes from the need to observe the two
celestial sources for 1 Ms each to achieve accuracies shown in Table 1. To
estimate mission duration, we divide $2\times10^6$\,s by 40\% (estimated
fraction of night time), by 80\% (estimated fraction outside SAA), by 50\%
(to allow observations of SrcSat) and arrive at $\sim 1.3\times 10^7$ s, or
about 5 months, which drives the primary mission length. A 6-month mission
extension allows internal checks or additional targets.

The two CubeSats will be deployed together and achieve the needed separation
using the TelSat propulsion system.

%%%%%%%%%%%%%%%%%%%%%%%%%%%%%%%%%%%%%%%%%%%%%
\begin{table*}[t]
\small
\noindent
\hspace*{-1mm}
\renewcommand{\tabcolsep}{3mm}
\renewcommand{\arraystretch}{1.2}
\begin{tabular}{ccllcc}
\multicolumn{6}{c}{\normalsize Table 2. Mission Parameters}\\
\hline
\hline
Mass (kg)  & Cube size (U) &  Desired orbit & Acceptable & ISS orbit OK? &  In-orbit life \\
\hline
24 + 12  & 12 + 6 & Altitude (km): 500 & 500--600 & No  & 1 yr \\
         &        & Inclination (deg): 28 & 28-55 &    & \\
\hline
\end{tabular}
\end{table*}
%%%%%%%%%%%%%%%%%%%%%%%%%%%%%%%%%%%%%%%%%%%%%

% %%%%%
% \begin{table*}
% \small
% \noindent
% \hspace*{-1mm}
% \renewcommand{\tabcolsep}{2mm}
% \renewcommand{\arraystretch}{1.2}
% \begin{tabular}{p{3.5cm}p{2.5cm}p{10cm}}
% \multicolumn{3}{c}{Table 3: Mission Accommodation Requirements}\\
% \hline
% \hline
% Parameter &      Requirement  &   Implementation \\
% Satellite separation  &  1--2 km & Single deployer, matched ballistic
% coefficients (20\%), propulsion (TelSat only) \\
% Separation knowledge & 6m ($3\sigma$) &  Absolute Positions obtained from GPS\\
% Source orientation & 3\deg\ & Point SrcSat towards RAM during calibration \\ 
% Telescope pointing & 0.5\am\ & FAS error signal provided to ACS; star
% trackers \\
% Daily data volume & $<200$ Mb / sat & S-band, NEN, one contact/day/satellite
% \\
% \hline
% \end{tabular}
% \end{table*}
% %%%%%

\section{ORGANIZATION AND PARTNERSHIPS} 

NASA Goddard will build most of the payload that is not off-the-shelf.  NIST
leads absolute calibration of the radioactive sources and MXS.
Collaborators from Open University (UK) lead the CCD effort. We have X-ray
calibration experts from Harvard-Smithsonian CfA. The CubeSat bus will be
procured from Blue Canyon Co., and we will collaborate with them on
integrating Goddard's camera with their navigation algorithms.  The
extendable boom requires collaboration with Orbital ATK Co. All these
collaborations are currently ongoing.

\section{TECHNOLOGY DRIVERS} 

The constancy of the 1.5 keV to 5.9 keV line ratio for the MXS remains to be
demonstrated. It is the subject of an ongoing internally-funded study at
Goddard. The absolute calibration of the 1.5 keV line produced by MXS to the
required accuracy is under current study at NIST; the current accuracy for
1.5 keV is lower than that for $^{55}$Fe, which is taken into account in our
estimates.

\section{SCHEDULE AND COST} 

We have performed a mission study in 2018 for an APRA CubeSat opportunity at
NASA Wallops Mission Planning Lab (MPL). The schedule, based on in-house
experience, allocates 3 years from start of funding to build the components
and integrate the two satellites, about 3/4 year waiting for the ride, and 1
year in orbit.

The MPL study also produced a cost estimate based on the master equipment
list, vendor quotes, in-house experience for X-ray mirror manufacturing and
development of navigation components (FAS), and consultations with NIST.
The cumulative cost over the 5-year mission cycle (including 1 year in
orbit) is \$12M, of which equipment is \$5M, labor \$5M, and the CubeSat
launch \$1M.  Implementing \calx\ as a SmallSat (rather than APRA) will
allow us to use higher-reliability bus and components, which is what we are
envisioning. It will require budget reserves, and the launch will be more
expensive. The total will increase but certainly stay under the SmallSat cap
of \$35M.

\clearpage
\section{REFERENCES}

%%%%%%%%%%%%%%%%%%%%%%% REFERENCES:
\vspace{4mm}
\parindent=0cm \parskip=0cm
\small \baselineskip=14pt
\def\refpar{\par\hangindent=1.2em\hangafter=1}
\def\reference#1{\par\hangindent=1.2em\hangafter=1}

\reference{} AlSat 2016, http://space.skyrocket.de/doc\_sdat/alsat-nano.htm

\reference{} Benson, B., et al. 2014, ``SPT-3G: a next-generation cosmic
microwave background polarization experiment on the South Pole telescope'',
Proc.\ SPIE, 9153, 91531P

\reference{} Boldt, E. and Leiter, D., 1995, ``The Cosmic X-ray Background
as a measure of history'', Nucl.\ Phys.\ B, 38, 440

\reference{} Gendreau, K. C. et al., 2012, ``The x-ray advanced concepts
testbed (XACT) sounding rocket payload'', Proc.\ SPIE, 8443

\reference{} Harris, R. D. et al. 2011, ``Compact CMOS Camera Demonstrator
(C3D) for Ukube-1'', Proc.\ SPIE, 8146

\reference{} \hitomi\ Collaboration, 2018, PASJ in press, arXiv:1802.05068

\reference{} Madsen, K., et al. 2017, ``IACHEC Cross-calibration of
\chandra, \nustar, \swift, \suzaku, \xmmnewt\ with 3C 273 and PKS
2155-304'', AJ, 153, 2

\reference{} Madsen, K., et al. 2019, ``IACHEC'', Project White Paper for
the Astro-2020 Decadal Survey

\reference{} Markevitch, M. 2007, ``Helium abundance in galaxy clusters an
Sunyaev-Zeldovich effect'', arXiv:0705.3289

\reference{} Matheson, H. and Safi-Harb, S. 2005, ``The Plerionic Supernova
Remnant G21.5-0.9: In and Out'', Adv. Sp. Res., 35, 1099.

\reference{} McEachen, M. E., ``Development of the GEMS Telescope Optical
Boom''. 52nd AIAA/ASME/AHS/ASC Structures, Structural Dynamics and Materials
Conference, 4-7 April 2011, Denver, CO

\reference{} McEachen, M. E., ``Verification \& Placement Precision and
stability for the GEMS Telescope Optical Boom'', 54th AIAA/ASME/AHS/ASC
Structures, Structural Dynamics and Materials Conference, 11-18 April 2013,
Boston, MA

\reference{} Merloni, A., et al. 2012, ``eROSITA Science Book: Mapping the
Structure of the Energetic Universe'', arXiv:1209.3114

\reference{} Ozel, F., et al. 2016, ``The Dense Matter Equation of State
from Neutron Star Radius and Mass Measurements'', ApJ, 820, 28

\reference{} Park, J-P et al. 2016, ``Mission Analysis and CubeSat Design
for CANYVAL-X mission,'' AIAA Space Ops Conference 2016, Daejeon, Korea

\reference{} Planck Collaboration, 2016, ``Planck 2015 results XXIV.
Cosmology from Sunyaev-Zeldovich cluster counts'', A\&A 594, A24

\reference{} Plucinsky, P. et al. 2018, in preparation,
http://web.mit.edu/iachec/meetings/2017/presentations/ IACHEC2017\_Plucinsky\_Thermal\_SNRs.pdf

\reference{} Pollock, A., and Guainazzi, M., 2014, ``Stability of the
XMM-Newton EPIC-pn and Variability of X-ray Sources'', in The X-ray Universe
2014, 302

\reference{} Razdolescu, A. C. et al., 2008, ``Measurement of $^{55}$Fe
solution activity by LSC-TDCR method'', Applied Radiation and Isotopes, 66,
750.

\reference{} Schellenberger, G. et al., 2015, ``XMM-Newton and Chandra
cross-calibration using HIFLUGCS galaxy clusters'', A\&A, 575, A30

\reference{} Silk, J., \& White, S.D.M. 1978, ``The determination of qo
using X-ray and microwave observations of galaxy clusters'', ApJ, 226,
L103

\reference{} Steiner, J.F. et al., 2014, ``The Low-Spin Black Hole in LMC
X-3'', ApJ, 793, 29.

\reference{} Tsujimoto, M., Guainazzi, M., Plucinsky, P.~P., et al. 2011,
``Cross-calibration of the X-ray instruments onboard the \chandra, INTEGRAL,
RXTE, Suzaku, Swift, and XMM-Newton observatories using G21.5-0.9'', A\&A,
525, A25.

\reference{} Vikhlinin, A., et al., 2009, ``\chandra\ Cluster Cosmology
Project III: Cosmological Parameter Constraints'', ApJ, 692, 1060.

\end{document}